\documentclass[pre,twocolumn,superscriptaddress,amsmath,amssymb,floatfix]{revtex4-2}
\usepackage{graphicx}
\usepackage{booktabs}
\usepackage[colorlinks=true,allcolors=blue]{hyperref}

\begin{document}

\title{Scale-rich network metamaterials grow by local rules on a recursive cell division}

\author{Alexei Vazquez}
\email{alexei@nodeslinks.com}
\affiliation{Nodes \& Links Ltd, Salisbury House, Station Road, Cambridge, CB1 2LA, UK}

\date{\today}

\begin{abstract}
Scale-rich (SR) metamaterials are planar ligament networks grown by nucleating a
segment at a random point and angle and extending it until it meets the existing
structure. In the dual graph, where ligaments are nodes and junctions are links, the
degree distribution has a $k^{-3}$ tail that was attributed to a spatial preferential
attachment. We show that the SR dual belongs to the class of local growth rules.
The construction is Cowan's iterated cell division; as a graph, each new node joins two
nodes on the boundary of one face. The degree distribution follows from a rate equation whose attachment kernel
$(a+bk)/t$ is deduced from this local sampling. A ligament of degree $k$ borders $k$
cells, and a random point reaches it through them, so $b=2/\langle n\rangle=1/2$ with
$\langle n\rangle=4$ sides per cell, and a sum rule gives $a=0$. The rate equation then
gives $p_k=12/[k(k+1)(k+2)]$ with no free parameter. The measured kernel and degree
distribution agree with this prediction. Preferential attachment is thus the mean-field outcome of
local sampling, not a mechanism. Ligaments are hit in proportion to their length,
which is fixed at birth, so the effective attachment carries no reinforcement. The same face-local step produces the signature of the local class.
The probability that a new ligament cuts off a corner of its cell sets the triangle
closure and the hierarchical clustering spectrum
$C(k)\simeq2p/k$, and square-closing splits raise the shortest-path multiplicity above
its value at fixed degree sequence. Thickness, jamming, angle sets and density gradients
leave $\gamma=3$; coherent orientation fields give $\gamma\simeq2.4$--$2.5$. In $d$
dimensions the cell hierarchy fixes $b=(d-1)/d$, so plates in three dimensions have
$b=2/3$ and $\gamma=5/2$, as measured. The
reported per-ligament energy law cannot distinguish degree from length or age.
\end{abstract}

\maketitle

\section{Introduction}

Both \textit{et al.}~\cite{both2026} introduced scale-rich (SR) mechanical
metamaterials, two-dimensional ligament networks in which length, thickness and
connectivity span orders of magnitude. The construction has two parameters. At step
$t=1,2,\dots,T$ a point is drawn uniformly in the unit square together with a uniform
angle, and a ligament of thickness $\lambda_t=\lambda_0t^{-\alpha}$ is grown both ways
from it until each end meets an existing ligament or the boundary. Every junction is a
T. Following the road-network convention~\cite{rosvall2005,kalapala2006}, the authors
represent each ligament as a node and each junction as a link. They report power-law
distributions of thickness, length and degree, with $P(k)\sim k^{-3}$ attributed to a
spatial preferential attachment, because longer ligaments offer more opportunities to
be hit. They also report a per-ligament elastic energy law
$U_i\sim k_i^{3/2}\lambda_i^{1/2}f(\theta_i)$ and a density scaling of the total
energy controlled by $\alpha$~\cite{both2026}.

Two questions are natural from the point of view of network growth. Which growth
class does the SR dual graph belong to, and which of its properties are fixed by
that class rather than by the design parameters $(\alpha,\lambda_0)$? Local growth
rules, in which a new node attaches to an existing node and to some of its
neighbours, produce a hierarchical clustering spectrum $C(k)\sim1/k$ and degree
correlations~\cite{vazquez2003,holme2002,dorogovtsev2001,ravasz2003}. They also
produce preferential attachment as an effective result: a link followed at random
reaches a node with probability proportional to its degree, and the rate equation
approach~\cite{krapivsky2000}, with a kernel deduced from the rule, gives the degree
distribution~\cite{vazquez2003}. The geometric construction is itself not new: with
zero thickness it is Cowan's iterated cell division~\cite{cowan2010}, a member of the
family of random tessellations that also contains the STIT
model~\cite{nagel2005,redenbach2025,*redenbach2025arxiv} and related to hierarchical crack patterns formed
by successive domain divisions~\cite{bohn2005,bohn2005b}. We show that the SR dual
belongs to the local class: each new node joins two nodes on the boundary of one face,
and this step produces both the clustering and the degree law. The attachment kernel
follows from the face-local sampling, $b=2/\langle n\rangle=1/2$, and the rate equation
approach~\cite{krapivsky2000} then gives the degree distribution exactly in mean field. The
spatial preferential attachment of Ref.~\cite{both2026} is this mean-field outcome, not
its explanation. Inside the face the geometry samples by length, assigned at birth by
the cell hierarchy, so the effective preferential attachment carries no reinforcement,
as the fluctuations of the degree at fixed length and age show. Keeping both references in view, the recursive division and the local graph
rule, turns most of the reported network statistics into exact or asymptotically exact
statements.

The paper is organised as follows. Section~\ref{sec:model} maps the SR construction
onto Cowan's division and derives the statistics of the cell hierarchy. In
Sec.~\ref{sec:degree} we write the dual growth as a face-local rule, deduce its
attachment kernel, solve the rate equation and test it against the growth history. Section~\ref{sec:triangles} computes the corner-cut probability
by integral geometry and derives the clustering spectrum. Section~\ref{sec:topology}
asks which properties survive when geometry is removed, and Sec.~\ref{sec:paths}
turns to distances and shortest-path multiplicity. Section~\ref{sec:gamma} identifies
which rules move the degree exponent, Sec.~\ref{sec:3d} treats plates in three
dimensions, and Sec.~\ref{sec:energy} discusses what this
means for the reported structure--property relations.

\begin{figure*}[tbp]
\includegraphics[width=\textwidth]{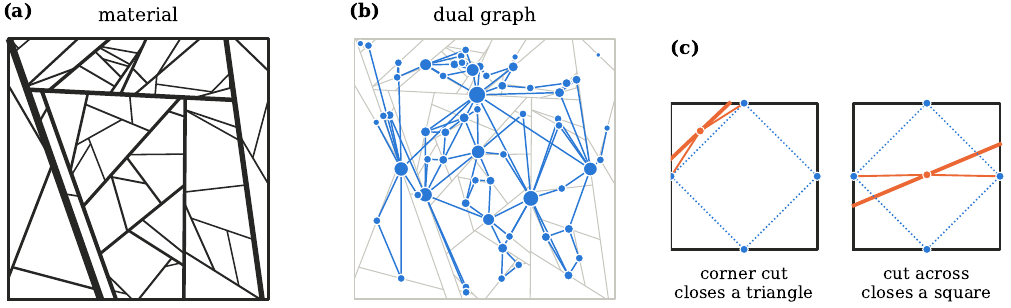}
\caption{(a) An SR material with $T=60$ ligaments. The line width increases with
$\lambda_t=\lambda_0t^{-1/2}$. (b) Its dual graph: each ligament is a node, placed
at the ligament midpoint with a size that grows with its degree, and each junction is
a link. (c) A new ligament (orange) splits one cell. Dual nodes of the cell's sides
are shown in blue, and the dotted lines join sides that meet at a corner. If the new
ligament cuts off a corner, it joins two adjacent sides and closes a triangle in the
dual. Otherwise it typically joins two sides at distance two and closes a square; in
cells with six or more sides the two sides can be farther apart, and $5\%$ of new
ligaments close a triangle with a ligament outside the cell.}
\label{fig:model}
\end{figure*}

\section{The SR construction is a recursive cell division}
\label{sec:model}

We first set $\lambda_0=0$ and return to thickness at the end of this section. The
current ligaments partition the square into convex polygonal cells. A uniform
point falls in cell $K$ with probability equal to its area $A_K$. Because $K$ is
convex and bounded by ligaments, the new ligament grown from that point stops on the
boundary of $K$: it is the chord of $K$ through a uniform point of $K$ at a uniform
angle. The SR model is therefore the iterated division of cells with
\emph{area-weighted selection} and the \emph{chord through a uniform point} division
rule, model~(a) of Fig.~2 in Cowan~\cite{cowan2010}. Cowan rescales the window after
each division, by a factor that keeps the mean cell size fixed, so that a stationary
limit exists; the rescaling is a similarity and does not change the topology. STIT
tessellations differ in both rules: cells are selected by perimeter and cut by
isotropic uniform random lines~\cite{nagel2005,redenbach2025,*redenbach2025arxiv}. Gilbert's
model~\cite{gilbert1967}, used as a reference architecture in Ref.~\cite{both2026},
grows all segments simultaneously and is a different process. Hierarchical crack
patterns in drying layers also divide domains one at a time and form T-junctions, but
the division is deterministic or nearly so, set by the domain shape and the layer
thickness~\cite{bohn2005,bohn2005b}. Throughout, the boundary of the square is not a
node of the dual graph; $0.55\%$ of ligament ends lie on it.

\paragraph*{Corner numbers.}
A junction placed on the side of a neighbouring cell is a straight angle there, so
the corner number $n$ of a cell is fixed at birth. A chord ending on two sides gives
each daughter the two new corners at its ends, so $n_1+n_2=n+4$. Starting from the
square, after $t$ steps there are $t+1$ cells and $4+4t$ corners, so the mean corner
number is exactly $\langle n\rangle=4$. Measured number fractions for $n=3,\dots,7$
are $0.301,0.452,0.200,0.042,0.005$. A Markov-chain treatment of the side number under
division, as done for proliferating epithelia~\cite{gibson2006}, would give these
fractions once the corner-cut probabilities of Sec.~\ref{sec:triangles} are known by
side number; the rules differ from the epithelial case because T-junctions do not add
corners to the neighbours.

\begin{figure*}[tbp]
\includegraphics[width=\textwidth]{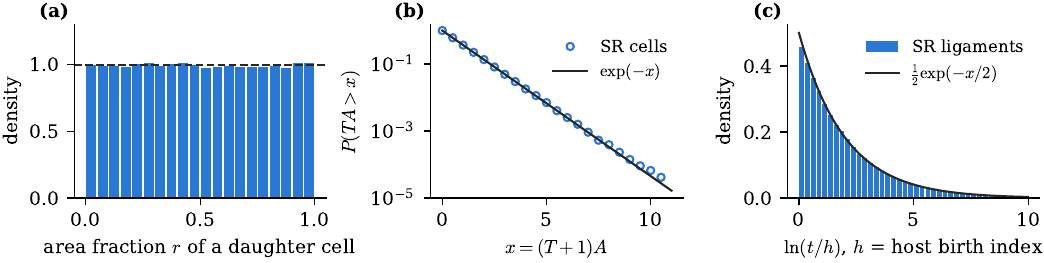}
\caption{The cell hierarchy of the SR construction (six realisations, $T=2\times10^4$).
(a) The area fraction of a daughter cell is uniform. (b) Rescaled cell areas follow
$\exp(-x)$, the spacing law of uniform points. (c) The birth index $h$ of the ligament
hit by a new ligament $t>2000$ follows the mean-field law $h/t\to U^2$, for which $\ln(t/h)$
has density $\frac12\exp(-x/2)$.}
\label{fig:bst}
\end{figure*}

\paragraph*{Uniform splitting and the binary search tree.}
Let $K$ be a convex body, $x$ a uniform point in $K$, and $\theta$ any direction
independent of $x$. For fixed $\theta$, the offset $s$ of the line through $x$ along
the normal direction has density $w(s)/A_K$, where $w(s)$ is the chord length at
offset $s$. The area on one side of the line is $\int_{-\infty}^s w(s')ds'$, so the
area fraction is the cumulative distribution function of $s$ evaluated at $s$. It is
therefore uniform on $(0,1)$, whatever the shape of $K$ and the law of $\theta$. This
uniformity and the resulting exponential law of rescaled areas are Cowan's
results~\cite{cowan2010}. They have a compact restatement. Label each cell by the
interval of $[0,1]$ that it would occupy if areas were laid on a line. With
area-weighted selection, the sequence of splits is exactly that of a binary search tree
(BST) built from uniform keys, the familiar random scission of an interval: the key
falls in an interval with probability equal to its length and splits it at the
key~\cite{devroye1986}. The cell areas after $T$ steps are therefore the $T+1$
spacings of $T$ uniform points, $P(TA>x)\to\exp(-x)$ and $\sum_KA_K^2=2/(T+2)$, and
the genealogy of cells inherits the BST statistics: the depth of the split at step $t$
grows as $2\ln t$, with variance $2\ln t$, and areas never multiscale. Our
simulations give a depth slope $1.983$, a variance slope $1.980$ and $\sum A^2=9.998\times10^{-5}$
at $T=2\times10^4$ against $9.999\times10^{-5}$ [Fig.~\ref{fig:bst}(a,b)]. The
multiscaling of Krapivsky and Ben-Naim~\cite{krapivsky1994} arises with rectangular
cuts and lives in the lengths, not in the areas. Axis-aligned chords reproduce their
width exponents $\xi(s)=(s-\sqrt{s^2+4})/2$ within $2\%$, while with isotropic
angles the perimeter moments of order $s$ scale as $t^{1-s/2}$, without multiscaling.

\paragraph*{Lengths and the age of the host.}
Cells at time $t$ have linear size $t^{-1/2}$, and the mean length of the ligament
born at $t$ is $\langle\ell_t\rangle\simeq c_\ell t^{-1/2}$ with $c_\ell=1.16$. The total
length grows as $L(t)\simeq c_L\sqrt t$ with $c_L=2c_\ell=2.33$. The first ligament has
$\langle\ell_1\rangle=0.945$, the $\ell_0$ of Ref.~\cite{both2026}. Section~\ref{sec:degree}
shows that an end of ligament $t$ lands on a given ligament with probability
proportional to its length. Replacing that length by its mean, a mean-field step, the
host's birth index $h$ satisfies $P(h)\simeq\ell_h/L(t)\simeq\tfrac12(ht)^{-1/2}$. Then
$P(h\le xt)=\sqrt x$, that is, asymptotically,
\begin{equation}
  h/t\;\to\;U^2,\qquad U\sim\mathrm{Uniform}(0,1).
  \label{eq:host}
\end{equation}
Equivalently, $\ln(t/h)$ is exponential with mean $2$ [Fig.~\ref{fig:bst}(c)].
Measured values of $\langle(h/t)^\alpha\rangle$ are $0.667$, $0.498$ and $0.330$ for
$\alpha=1/4$, $1/2$ and $1$, against $1/(1+2\alpha)=0.667$, $0.500$ and $0.333$.

\paragraph*{Thickness.}
With $\lambda_t=\lambda_0t^{-\alpha}$, Eq.~\eqref{eq:host} gives the thickness ratio
between a new ligament and its host as $\lambda_t/\lambda_h=U^{2\alpha}$, again
asymptotically and in mean field. Its law does not depend on $t$, so every generation
couples to the frame it attaches to in the same statistical way.

The fill fraction without overlaps,
$\rho\simeq\lambda_0\sum_t\langle\ell_t\rangle t^{-\alpha}$, converges only for
$\alpha>1/2$. This recovers the $\alpha=1/2$ line of the phase diagram of
Ref.~\cite{both2026}, but not its second branch: for $\alpha>1/2$ the structure also
jams when $\lambda_0>[\ell_0\zeta(2\alpha)]^{-1}$~\cite{both2026}. That branch comes from
the depletion of free area, which shortens later ligaments and which the overlap-free
estimate ignores. Including depletion in mean field, a ligament nucleated in free area
$F$ has length $\ell_t=c(t)(F/t)^{1/2}$, so
$dF/dt=-\lambda_0c(t)F^{1/2}t^{-\alpha-1/2}$ and
$\sqrt{F_\infty}=1-\tfrac{\lambda_0}{2}\sum_tc(t)t^{-\alpha-1/2}$, the two-dimensional
analogue of the recursion of Ref.~\cite{both2026} and of Eq.~\eqref{eq:F3d} below. The
structure jams when $\lambda_0\ge\lambda_0^*=2/\sum_tc(t)t^{-\alpha-1/2}$. With $c(t)$
measured in zero-thickness runs ($c\to c_\ell=1.163$), this gives $\lambda_0^*=0.168$,
$0.404$, $0.745$ and $1.254$ for $\alpha=0.6$, $0.75$, $1$ and $1.5$.

In finite-width simulations, where ligaments nucleate only in free area and stop at
the surface of the ligaments they meet, a run counts as jammed when the free area is
exhausted (below $10^{-14}$) before step $T$; $\lambda_0^*$ is bisected with two
realisations per value. At $T=2\times10^4$ we find
$\lambda_0^*=0.232$, $0.363$, $0.643$ and $0.958$. Near $\alpha=1/2$ the free-area sum
converges slowly, its remainder beyond $T$ scaling as $T^{1/2-\alpha}$, and the threshold
drifts: at $\alpha=0.6$ it falls to $0.214$ at $T=10^5$, while at $\alpha=0.75$ it is
stable ($0.359$). The formula of Ref.~\cite{both2026} gives $0.189$, $0.405$, $0.643$ and
$0.879$. The depletion recursion lies below the drifting threshold at $\alpha=0.6$, is within
$11\%$ at $\alpha=0.75$, and overestimates the
threshold by $16$--$31\%$ at $\alpha\ge1$, where jamming is decided by the first few
thick ligaments. The overlap-free estimate is too low by a factor of about $1.7$. Thickness removes area
and stops growth in the jammed phase, but it does not change the rule by which cells
are split (Sec.~\ref{sec:gamma}). In what follows the network is the zero-thickness
one unless stated otherwise.

\section{A face-local growth rule and its rate equation}
\label{sec:degree}

In the dual graph, step $t$ reads as follows: choose a face (a cell) with probability
equal to its area, choose the two sides of that face that the chord meets, and join
the new node $t$ to the two ligaments that carry them [Fig.~\ref{fig:model}(c)]. The new
node never links to a node outside the chosen face. It is a local rule in the sense of
Refs.~\cite{vazquez2003,vazquezbook}, with faces in place of neighbourhoods, and its
degree distribution follows from a rate equation once the rate at which a node of
degree $k$ is hit is deduced from the rule.

\paragraph*{Incidences equal degree.}
Call an incidence a pair (ligament $i$, cell $K$) in which $i$ contributes a side to
$K$. A new ligament has two incidences, one on each side, and two links. When a later
ligament ends on $i$, it adds one link to $i$ and cuts one of $i$'s sides in two, adding
one incidence. For a ligament whose ends are both interior, the number of incidences
therefore equals the degree $k_i$ at all times: a ligament of degree $k$ borders $k$
cells. Each step creates two hits and four incidences, so the total number of
incidences is $4t$, in agreement with $\langle n\rangle=4$. This identity is exact.

\paragraph*{The attachment rate.}
A uniform point selects a cell, and the chord ends
on the cell's boundary. Ligament $i$ is therefore reached through the cells it borders,
and the rate can be counted in two ways.

(i) \emph{By incidences.} Ligament $i$ borders $k_i$ cells. If these are typical cells,
each selected with probability $1/t$ on average, and each of a cell's $\langle n\rangle$
sides receives one of the two ends with equal probability, $i$ is hit at rate
$k_i\cdot(1/t)\cdot(2/\langle n\rangle)=k_i/2t$. The more cells a ligament borders, the
more often a random point reaches it, and it borders one cell per link. The cells next
to hubs are indeed typical. In units of the mean, their mean area is $1.04$ at $k=2$ and
$1.15$--$1.17$ for $10\le k<160$ ($T=10^5$, three realisations): hubs border many typical
voids, not large ones.

(ii) \emph{By length.} A chord end lands on the structure at a point that is, in mean
field, uniform per unit length. Two ends land per step on a total length
$L(t)=c_L\sqrt t$, so ligament $i$ is hit at rate $2\ell_i/L(t)$, and integrating,
\begin{equation}
  k_i-2=\frac{4\ell_i}{c_L}\left(\sqrt t-\sqrt i\right)=2x_i\left(\sqrt{t/i}-1\right),
  \label{eq:kl}
\end{equation}
where $x_i=\ell_i/\langle\ell_i\rangle$ and $4c_\ell/c_L=2$. This is the ``more
opportunities to be hit'' argument of Ref.~\cite{both2026}, made quantitative.
Differentiating and eliminating $\sqrt{t/i}$ gives $dk_i/dt=(k_i-2+2x_i)/2t$. In mean
field, $x_i\to1$, this is $k_i/2t$ again: the ligaments that have collected many links are
the long ones.

Both counts give an attachment rate linear in the degree, $(a+bk)/t$ with $a=0$ and
$b=2/\langle n\rangle=1/2$, exact within mean field because $\langle n\rangle=4$ is exact.
Preferential attachment is thus the mean-field outcome of local
sampling~\cite{vazquez2003,vazquezbook}. The spatial
preferential attachment of Ref.~\cite{both2026} is count (ii).

\paragraph*{Rate equation.}
Following the rate equation approach~\cite{krapivsky2000}, let $n_k(t)$ be the
expected number of nodes of degree $k$ after $t$ steps and $q_k$ the probability that
the newcomer makes $k$ links at birth (in 2D $q_k=\delta_{k,2}$). Then
\begin{equation}
  n_k(t+1)=n_k(t)+\frac{a+b(k-1)}{t}\,n_{k-1}-\frac{a+bk}{t}\,n_k+q_k .
  \label{eq:rate}
\end{equation}
In the stationary regime $n_k=t\,p_k$, and
\begin{equation}
  p_k=\frac{[a+b(k-1)]\,p_{k-1}+q_k}{1+a+bk}.
  \label{eq:rec}
\end{equation}
For a single birth degree $o$ this unrolls into
$p_k=b^{-1}\,\Gamma(o+\nu)\Gamma(k+\eta)/[\Gamma(o+\eta)\Gamma(k+1+\nu)]$ with
$\eta=a/b$ and $\nu=(1+a)/b$, a power law with
\begin{equation}
  \gamma=1+\frac1b .
  \label{eq:gammab}
\end{equation}
The constants are tied by a sum rule. Every hit adds one link, so the rates sum to the
mean number $\langle o\rangle$ of hits per step, and $\langle k\rangle=2\langle o\rangle$;
hence $a+2b\langle o\rangle=\langle o\rangle$, that is
\begin{equation}
  a=\langle o\rangle(1-2b),
  \label{eq:sumrule}
\end{equation}
which $a=0$, $b=1/2$ and $\langle o\rangle=2$ satisfy. With $o=2$, $\eta=0$ and $\nu=2$,
\begin{equation}
  p_k=\frac{12}{k(k+1)(k+2)},\qquad P(k'\ge k)=\frac{6}{k(k+1)},
  \label{eq:pk}
\end{equation}
and $\gamma=3$.

\begin{figure}[t]
\includegraphics[width=\columnwidth]{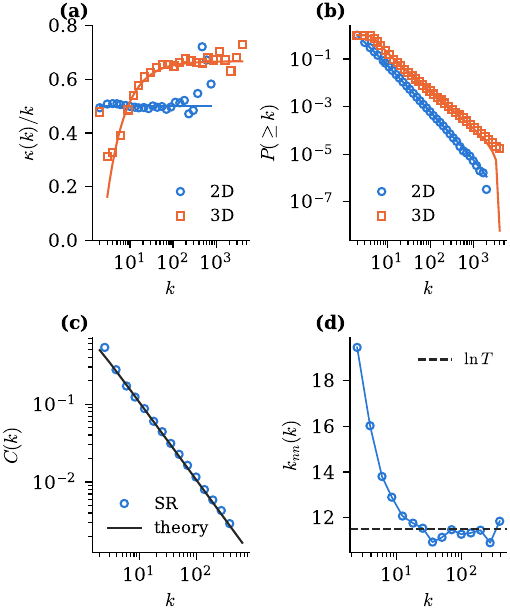}
\caption{Degrees and clustering of the SR dual. (a) Attachment kernel measured from the
growth history, $\kappa(k)/k$ with $\kappa=t\,r_k$, in two dimensions ($T=10^6$, three
realisations) and for plates in three dimensions ($T=2\times10^6$, three realisations).
Lines: deduced kernels, $\kappa=k/2$ and $\kappa=-\langle o\rangle/3+2k/3$
(Secs.~\ref{sec:degree} and \ref{sec:3d}). (b) Degree distributions (symbols) and the
stationary solution of the rate equation~\eqref{eq:rec} with the deduced constants and
the measured birth degrees (lines); no parameter is fitted. (c) Clustering spectrum,
$T=10^5$, and Eq.~\eqref{eq:ck} with $p_0=0.494$, $p=0.53$. (d) Mean neighbour degree,
$T=10^5$; the dashed line is $\ln T$.}
\label{fig:degree}
\end{figure}

\paragraph*{Test against the growth history.}
We measured the kernel directly. With $H_k$ the number of hits on nodes of degree $k$
during $t\in[T/10,T]$ and $n_k(t)$ tracked exactly at every step,
$\kappa(k)=H_k/\sum_tn_k(t)/t$ estimates $t\,r_k=a+bk$, where $r_k$ is the rate at
which a node of degree $k$ is hit. At $T=10^6$ (three
realisations) a linear fit for $k\le200$ gives
$\kappa=-0.006+0.502\,k$ (intercepts $-0.002$ to $-0.008$ over realisations) [Fig.~\ref{fig:degree}(a)].
The stationary solution of Eq.~\eqref{eq:rec} with the deduced constants, and no free
parameter, matches the simulated distribution [Fig.~\ref{fig:degree}(b)]. The cohort born at $i$ has mean
degree $\propto i^{-\beta}$ with $\beta=0.50$--$0.51$, as $k_i\propto t^b$ requires.

\paragraph*{Beyond mean field: which ligament is hit.}
The rate equation fixes the population. Count (ii) also describes the individual
ligament: it is hit at a rate set by its birth length, $x_i/\sqrt{it}$, and not by the
links it has accumulated. Being hit splits a side but adds no length. Equation~\eqref{eq:kl}
predicts the degree of each ligament with mean ratio $1.006$ and correlation $0.95$, and
the deviations behave as independent rare events. (1) In bins of the predicted mean
$\mu_i\in[2,4),[4,8),\dots,[64,128)$ the dispersion
$D=\langle(k_i-2-\mu_i)^2\rangle/\langle\mu_i\rangle$ is $0.92$--$0.94$, $0.93$--$0.96$,
$1.01$--$1.03$, $1.07$--$1.12$, $1.05$--$1.38$ and $0.92$--$1.06$ ($T=10^5$, three
realisations). (2) Hits during $[i,2i)$ and during $[2i,T)$, each taken relative to
Eq.~\eqref{eq:kl}, are uncorrelated, overall ($r=-0.027$, $-0.015$, $0.045$) and for the
oldest ligaments with $\mu_i>16$ ($r=-0.017$, $0.032$, $0.094$). The topological surrogate
of Sec.~\ref{sec:topology} selects sides uniformly and has the same kernel, $b=1/2$, by
construction. There, each hit adds an incidence and raises the future rate, and the
same tests give a variance $10$--$91$ times the mean at fixed age and an early--late
correlation of $0.46$--$0.50$. The geometry thus realises the mean-field kernel without
reinforcement. The length weighting leaves $b$ and $\gamma$ unchanged but sets the tail
amplitude. At $T=10^6$ (five realisations) $K^2P(k\ge K)=5.78\pm0.02$, $5.75\pm0.03$ and
$5.75\pm0.06$ at $K=32$, $64$ and $128$, about $3\%$ below the values $6K/(K+1)=5.82$,
$5.91$ and $5.95$ of Eq.~\eqref{eq:pk}. Drawing each degree as $2$ plus a Poisson variable
with the mean of Eq.~\eqref{eq:kl}, multiplied by the measured ratio of received to
predicted hits ($1.01$--$1.025$ for $\mu_i>2$; old ligaments receive about $2\%$ more
than the mean-field rate $2/L(t)$), gives $5.70$--$5.78$, which reproduces the data.

\section{Corner cuts, triangles and clustering}
\label{sec:triangles}

A new node closes a triangle in the dual when the two ligaments it joins are already
linked. Locally this happens when the chord cuts off a single corner of its cell,
since the two sides meeting at that corner belong to linked ligaments.

\paragraph*{Integral-geometry formula.}
Chords through a uniform point at a uniform angle sample lines with density
$w(s)\,ds\,d\theta/(\pi A_K)$. For a fixed direction $\theta$, order the corners by
their projection on the normal. The chord cuts off the lowest corner alone when its
offset lies between the projections of the lowest and second-lowest corners. The
integral of $w(s)$ over that range is the area $A_{\rm lo}(\theta)$ cut off by the line
through the second-lowest corner, and likewise $A_{\rm hi}(\theta)$ at the other end.
Hence
\begin{equation}
  P_{\rm corner}(K)=\frac{1}{\pi A_K}\int_0^\pi\big[A_{\rm lo}(\theta)+A_{\rm hi}(\theta)\big]\,d\theta .
  \label{eq:corner}
\end{equation}
The line measure is the standard one of integral geometry~\cite{santalo1976}, weighted
here by the chord length because the line passes through a uniform point. For a
triangle every chord cuts off a corner, so $P_{\rm corner}=1$. For a square,
Eq.~\eqref{eq:corner} gives $2\ln2/\pi=0.4413$ in closed form. Numerical quadrature of
Eq.~\eqref{eq:corner} gives $0.303$ for a $1\times4$ rectangle and $0.232$ and $0.136$ for
regular pentagons and hexagons. Evaluated on the simulated SR cells, the mean of
Eq.~\eqref{eq:corner} by corner number $n=3,\dots,7$ is $1.000$, $0.461$, $0.254$,
$0.152$ and $0.10$, and its area-weighted average is $0.447$--$0.449$ (numerical). The
frequency of corner cuts measured directly in the growth is $0.447$--$0.449$. This
agreement checks the formula and the code against each other; it is not an
independent prediction, because the cell shapes are taken from the simulation.

\paragraph*{Triangle bookkeeping.}
A corner cut creates a triangular daughter unless the parent was already a
triangle, which can only split into $(3,4)$. The number fraction of triangles is
therefore $f_3=\langle P_{\rm corner}\rangle-w_3$, where $w_3$ is the area fraction of
triangles. Measured values are $0.30=0.449-0.147$.

\paragraph*{Closures through a link off the face.}
The fraction of new nodes that close a triangle is $0.496$, not $1/2$. Corner cuts
account for $0.448$. The other $0.048$ join two non-adjacent sides of their cell whose
ligaments already meet outside it. These steps are as local as the corner cuts: both
targets lie on the chosen face and are already linked, so the new node closes a
triangle on an existing link; only that link is not a side of the face. The remaining $0.50$ of new nodes join two ligaments
at distance two or more around the cell: distance two closes a square, and in cells
with $n\ge6$ the two sides can be three steps apart around the face.

\paragraph*{Clustering spectrum.}
Links between existing nodes are never added. A node therefore acquires triangles only
at birth, with probability $p_0$, or when it is hit and the other end of the new
ligament lands on one of its neighbours, with probability $p$ per hit. With constant
$p$ the number of triangles at node $i$ is $e_i=p_0+p(k_i-2)$, and
\begin{equation}
  C(k)=\frac{2[p_0+p(k-2)]}{k(k-1)}\simeq\frac{2p}{k},
  \label{eq:ck}
\end{equation}
the hierarchical form produced by triadic
closure~\cite{vazquez2003,ravasz2003,holme2002}. Counting the per-hit triangle events
directly, $p$ is not constant: it is $0.41$ when the hit node has degree $2$, $0.48$ for
degrees $3$--$4$, $0.53$ for $5$--$9$, and $0.56$--$0.58$ for degrees $10$ and above
(two realisations, $T=10^5$, each bin $\ge10^4$ hits). Hits on hubs are more likely to
land next to a neighbour because hubs are long and carry many short stems. With
$p_0=0.494$ and a single fitted $p=0.53$ ($0.56$ when the fit is restricted to $k>7$),
Eq.~\eqref{eq:ck} is within $2\%$ of the data for $k\ge15$ [Fig.~\ref{fig:degree}(c)].
The mean clustering is $0.38$. At $k=2$ the measured $C=0.536$ exceeds $p_0$, because
corner-cut ligaments are short and rarely hit again. The degree assortativity is
$r\simeq-0.02$. The mean neighbour degree $k_{nn}(k)$ falls from $19.5$ at $k=2$ to a
plateau of $11.1$--$11.5$ for hubs, close to the value $\ln T=11.5$ expected for
a linear kernel with $b=1/2$ and birth degree $o=2$ [Fig.~\ref{fig:degree}(d)].

In the language of local rules, the SR dual mixes two moves, both confined to one
face. With probability $\simeq0.50$ the new node attaches to both ends of a link, as in
the edge-attachment model of Dorogovtsev, Mendes and Samukhin~\cite{dorogovtsev2001},
except that the link is selected through the area of a cell rather than uniformly; in
$0.45$ the link is a corner of the face and in $0.05$ it joins two of its sides outside
it. With probability $\simeq0.50$ it attaches to two nodes at distance two or more on a
face cycle, mostly closing a square. The edge-attachment model has $\gamma=3$ and
clustering $0.74$~\cite{dorogovtsev2001,dorogovtsevgoltsev2002}. The SR dual has the
same $\gamma$ and half the clustering; the other half of the closures went into squares.
It is also the two-dimensional relative of the random Apollonian
network~\cite{andrade2005,zhou2005}, in which faces are selected and a node's degree is
proportional to the number of faces it bounds.

\section{What topology alone reproduces}
\label{sec:topology}

To separate the role of geometry, we grew a purely topological surrogate. Each face is
stored as a cyclic list of the nodes that carry its sides. At each step a face is chosen
uniformly. If it is a triangle, two of its sides are joined; otherwise two adjacent
sides are joined with probability $p_4$, or two non-adjacent sides with probability
$1-p_4$. The face is then split into two, with $n_1+n_2=n+4$ as in the tessellation.
No lengths or positions are used.

Because every side of the chosen face is equally likely to be hit, every incidence is
equally likely to be hit. The surrogate therefore realises count (i) of
Sec.~\ref{sec:degree} exactly, with the same kernel, $b=1/2$. Every face-local variant we tried (faces chosen
uniformly, by number of sides, or by a random-split area) reproduces
Eq.~\eqref{eq:pk}: $\gamma_{\rm MLE}(k\ge10)=2.88$--$2.91$ and the same $K^2P$
amplitudes at $K\le32$. The single parameter $p_4=0.29$ also gives a mean clustering of
$0.386$ (SR, $0.377$) and places $C(k)$ and $k_{nn}(k)$ on the SR curves. The surrogate
reaches the kernel through reinforcement, each hit adding an incidence, and the
geometric model through length (Sec.~\ref{sec:degree}). Both are local rules with the
same rate equation: $P(k)$, $C(k)$ and $k_{nn}(k)$ are fixed by face locality, not by how
the sides are sampled.

The surrogate differs on distances. At $T=10^5$ its mean distance is $14.1$ against
$8.69$ for SR. Both grow logarithmically, but with prefactors $1.40$ and $0.60$
[Fig.~\ref{fig:paths}(a)]. In the geometric model chords hit long sides
preferentially, and long sides belong to old, long ligaments, so new nodes attach
close to the hubs. In the surrogate, sides are hit regardless of the age of the
ligament that carries them, and repeated cuts across faces can produce long chains of
squares. Geometry thus matters for the properties that depend on which particular sides
are hit, such as distances, and not for $P(k)$, $C(k)$ or $k_{nn}(k)$.

\begin{figure*}[tbp]
\includegraphics[width=\textwidth]{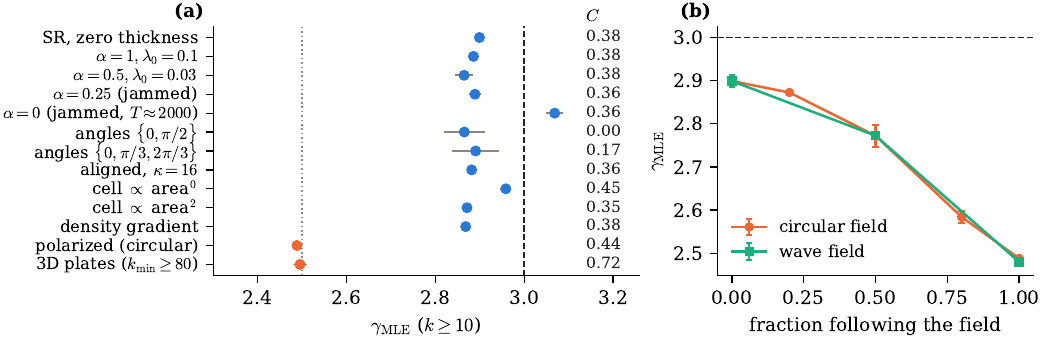}
\caption{(a) Maximum-likelihood exponent ($k\ge10$) for the design rules of
Ref.~\cite{both2026}, $T=3\times10^4$, two realisations (the jammed $\alpha=0$ case stops at
$T\simeq2000$; 3D plates: $T=2\times10^6$, three realisations, $k_{\min}=80$--$640$). The number on the right is the mean
clustering coefficient. Error bars: half the difference between the two realisations
(3D: range over $k_{\min}$). Dashed line: $\gamma=3$; dotted line: $\gamma=2.5$. The
uniform SR control gives $2.87$--$2.91$ at this size, the downward bias of the estimator
discussed in Sec.~\ref{sec:gamma}.
(b) Exponent against the fraction of ligaments whose angle follows a circular or a wave
orientation field ($T=3\times10^4$, two realisations; error bars: half their difference);
the remaining ligaments have uniform angles.}
\label{fig:gamma}
\end{figure*}

\section{Fewest-turn paths and their multiplicity}
\label{sec:paths}

A shortest path in the dual graph is a route through the material that changes
ligament the fewest times. For a random graph with given degrees, the mean number of
shortest paths between two nodes is $\langle\mu\rangle\simeq c_2/e$ with
$c_2=\langle k(k-1)\rangle/\langle k\rangle$, and local growth rules depart from this
value in a rule-dependent way~\cite{vazquezpaths2026}. For $\gamma=3$, $c_2$ grows as
$\ln T$.

We compare the SR dual with its degree-preserving randomisation (SR$^*$) and with the
edge-attachment (DM) model~\cite{dorogovtsev2001}, averaging over six realisations and
$100$ sources per realisation, for $T=10^3$ to $10^5$; over this range $c_2$ grows from
$7.9$ to $14.1$ and the mean distance of the SR dual from $5.78$ to $8.69$ ($4.48$ to
$6.06$ for SR$^*$). The randomised graph
sits on the configuration-model value, $\langle\mu\rangle/(c_2/e)=1.00$--$0.90$. The SR
dual exceeds it, by a factor that grows from $2.0\pm0.3$ at $T=10^3$ to $6.4\pm0.9$ at
$T=10^5$, and the DM model stays within $0.87$--$1.20$. Because $\mu$ has a heavy tail,
$\langle\mu\rangle$ is dominated by rare pairs; the mean logarithm is better determined.
The excess $\Delta\langle\ln\mu\rangle=\langle\ln\mu\rangle_{\rm SR}-\langle\ln\mu\rangle_{\rm SR^*}$,
taken at the same degree sequence, grows from $0.40$ to $1.03$ over the same range.

\begin{figure}[t]
\includegraphics[width=\columnwidth]{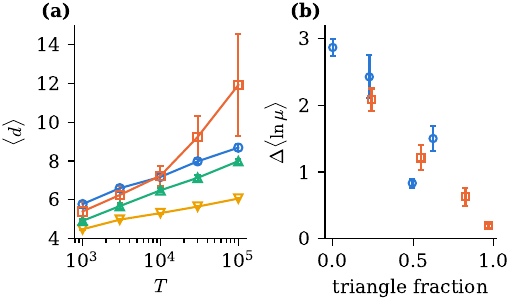}
\caption{(a) Mean distance in the dual graph against $T$ for the SR construction
(circles), its degree-preserving randomisation (down triangles), the edge-attachment (DM)
model~\cite{dorogovtsev2001} (up triangles), six realisations, and the topological face
surrogate with $p_4=0.29$ (squares), two to three realisations. (b) Excess of the mean log-multiplicity over each network's
degree-preserving randomisation against the fraction of new nodes that close a
triangle, $T=3\times10^4$, six realisations: SR rule variants (circles) (angle sets
$\{0,\pi/2\}$, $\{0,\pi/3,2\pi/3\}$, uniform angles, and uniform cell selection $q=0$) and
the face surrogate with $p_4=0.1$, $0.29$, $0.6$, $0.9$ (squares). Error bars: standard errors.}
\label{fig:paths}
\end{figure}

To ask where the excess comes from, we varied the fraction of new nodes that close a
triangle at fixed construction and measured $\Delta\langle\ln\mu\rangle$ against each
network's own randomisation [$T=3\times10^4$, six realisations,
Fig.~\ref{fig:paths}(b)]. In the topological surrogate, raising the corner-cut
probability $p_4$ from $0.1$ to $0.9$ raises the triangle-closing fraction from $0.24$
to $0.97$ and lowers the excess monotonically, from $2.09\pm0.17$ to $0.19\pm0.03$. In the
geometric model, the angle set $\{0,\pi/2\}$ closes no triangles and gives an excess
of $2.87\pm0.13$; the set $\{0,\pi/3,2\pi/3\}$ closes $0.23$ and gives $2.43\pm0.32$;
uniform angles close $0.49$ and give $0.83\pm0.07$. Square-closing splits are thus necessary for the redundancy, but their number alone
does not set it; their arrangement matters:
selecting cells uniformly ($q=0$) raises the triangle fraction to $0.62$ but also
raises the excess, to $1.50\pm0.19$. Uniform selection splits small cells as often as
large ones and stacks squares locally, which the area weighting of the SR rule
avoids. The DM model, which closes only triangles, has no excess.

\section{What moves the degree exponent}
\label{sec:gamma}

By Eq.~\eqref{eq:gammab} a rule moves the exponent only if it moves $b$, the growth
exponent of an old node, $k_i\propto t^b$. Count (ii) of Sec.~\ref{sec:degree}
deduces $b$ from two scalings: how the hit rate depends on length, and how the birth
length depends on the birth index. If the hit rate is proportional to $\ell_i^\sigma$
and $\ell_i\propto i^{-1/2}$, the total rate $\sum_j\ell_j^\sigma$ grows as
$t^{1-\sigma/2}$ for $\sigma<2$, so $k_i\propto\ell_i^\sigma t^{\sigma/2}\propto(t/i)^{\sigma/2}$,
giving $b=\beta=\sigma/2$ and $\gamma=1+2/\sigma$. In two dimensions with uniform angles
$\sigma=1$, so $\gamma=3$. Any rule that keeps the hit rate proportional to length and
the birth length proportional to the local cell size leaves $\gamma$ at $3$.

Exponents are estimated with the discrete maximum-likelihood approximation
$\hat\gamma=1+n[\sum_i\ln(k_i/(k_{\min}-1/2))]^{-1}$ over the $n$ nodes with
$k\ge k_{\min}$~\cite{clauset2009}, with standard error $(\hat\gamma-1)/\sqrt n$. For a
distribution with the curvature of Eq.~\eqref{eq:pk} this estimator is biased low at
small $k_{\min}$: on $10^5$ samples drawn exactly from Eq.~\eqref{eq:pk} it returns
$2.77\pm0.02$, $2.87\pm0.03$, $2.91\pm0.05$ and $2.91\pm0.10$ for $k_{\min}=6$, $10$, $20$
and $40$. The SR model at $T=10^5$ gives $2.77$--$2.79$, $2.86$--$2.90$, $2.94$--$2.97$
and $2.90$--$3.03$. Statements that $\gamma=3$ therefore rest on comparison with these
controls, not on the raw value of $\hat\gamma$.

We tested the design rules of Ref.~\cite{both2026} at $T=3\times10^4$, two
realisations each [Fig.~\ref{fig:gamma}(a)]:
\begin{itemize}
\item Thickness in the tunable-density phase ($\alpha=0.5$--$1$, $\lambda_0=0.01$--$0.1$,
$\rho=0.12$--$0.45$) gives $\gamma_{\rm MLE}=2.85$--$2.92$, $\beta=0.50$--$0.52$,
$C=0.376$--$0.378$, and a triangle closure of $0.49$--$0.50$.
\item In the jammed phase ($\alpha=0$--$0.4$), growth stops when the free area is
exhausted (at $t\simeq2000$ for $\alpha=0$, $\lambda_0=0.02$). Jamming truncates $T$; it
does not change $\gamma$ ($2.89$--$2.94$ wherever the statistics allow) or $C$
($0.35$--$0.37$).
\item Angle sets change clustering but not $\gamma$. With angles $\{0,\pi/2\}$ all cells are
rectangles, so $C=0$ exactly and $\gamma=2.87$; this is the rectangular fragmentation of
Ref.~\cite{krapivsky1994}. With $\{0,\pi/3,2\pi/3\}$, $C=0.17$.
\item Global alignment (von Mises, $\kappa\le16$), cell selection proportional to
area$^q$ ($q=0,0.5,2$), and density gradients give $\gamma=2.86$--$2.96$; $q=0$ and $q=2$
give $C=0.45$ and $0.35$.
\end{itemize}
The angle set and the selection weight tune clustering between $0$ and $0.45$; the
exponent stays at $3$.

\paragraph*{Coherent orientation fields.}
The polarized rule of Ref.~\cite{both2026}, in which the angle follows a prescribed
field $\theta(x,y)$, breaks the proportionality between hit rate and length. Locally
parallel newcomers cannot hit each other; they end on the ligaments that are misaligned
with the local field, and for straight ligaments in a curved field the misalignment
grows with length. The degree then grows faster than length, and birth lengths decay more slowly than
$i^{-1/2}$: $\langle\ell_i\rangle\propto i^{-0.35}$ for the circular field and $i^{-0.39}$
for the shortest wave, against $i^{-0.50}$ with uniform angles. Fitting
$\ln(k-2)$ against $\ln\ell$ for the oldest $5\%$ of ligaments with $k>4$ gives slopes
$\sigma=1.43$--$1.63$, against $1.03$--$1.05$ for uniform angles. With $\sigma\simeq1.5$ the
argument above with $\ell_i\propto i^{-1/2}$ predicts $\beta\simeq0.75$ and
$\gamma\simeq2.3$. With the measured birth-length decay $i^{-\delta}$ the same argument
gives $\beta=\sigma\delta\simeq0.53$--$0.6$ and $\gamma\simeq2.7$--$2.9$, now on the other side
of the measured values. Under fields the length at fixed birth time is broadly
distributed, and the exponent is set by the joint distribution of length and age rather
than by the two mean scalings; we have no theory for it. For a circular field and for waves $\theta=\pi(x+y)/w$ with
$w=0.125$--$1$ we find $\gamma\simeq2.44$--$2.51$ and cohort exponents $\beta\simeq0.6$--$0.67$,
stable from $T=10^4$ to $2\times10^5$. At $T=10^6$ the maximum-likelihood exponent is
$2.43$--$2.50$ ($k\ge10$ to $k\ge40$) for the circular field and $2.36$--$2.47$ for the
shortest wave ($w=0.125$), while the uniform control gives $2.88$--$2.98$, so the effect
does not wash out at this size. The exponent decreases continuously as the fraction of
ligaments following the field increases [Fig.~\ref{fig:gamma}(b)].

\section{Three dimensions}
\label{sec:3d}
With plates, the SR construction is the iterated division of convex polyhedra. A cell
is selected with probability equal to its volume, and it is cut by the plane through a
uniform point with an isotropic normal. The argument of Sec.~\ref{sec:model} carries over
unchanged: for a fixed normal, the offset of the plane has density $w(s)/V_K$, with $w(s)$
the section area, so the volume fraction of a daughter is uniform. The cell hierarchy is
again a binary search tree, with exponential volumes and split depth $2\ln t$. We measure
a daughter-fraction variance of $0.0833$ ($1/12$) and a depth slope of $2.01$. A plane that
crosses $m$ faces of a simple polyhedron changes $(F,V,E)$ by $(m+2,2m,3m)$ summed over the
two daughters. Hence the mean numbers of faces, vertices and edges per cell are
$2+\bar m$, $2\bar m$ and $3\bar m$, the counterpart of $\langle n\rangle=4$ (measured
$6.585$, $9.170$, $13.755$). Unlike in two dimensions, the number $m$ of plates met by a new
plate is not fixed by topology. Its mean converges to $\bar m=4.577$ for plates away from
the walls. Plate areas decay as $S_t\simeq c_St^{-2/3}$ with $c_S=1.23$, so the total area
grows as $3c_St^{1/3}$ and the area distribution is $P(S)\sim S^{-5/2}$ (maximum-likelihood
exponent $2.49$--$2.51$). The first plate has $\langle S_1\rangle=0.941$. Including plates
that meet the walls, the mean number of links made at birth is $\langle o\rangle=4.538$.

A plate is again reached through the cells it borders, and the two counts of the
attachment rate in Sec.~\ref{sec:degree} now give different shapes.

(i) \emph{By incidences.} A new plate creates $2+\langle o\rangle$ incidences and hands out
$\langle o\rangle$ links. If every incidence were equally likely to be hit,
$b=\langle o\rangle/(2+\langle o\rangle)=0.694$ and $\gamma=2+2/\langle o\rangle=2.44$.

(ii) \emph{By area.} A plate is hit at a rate proportional to its area, which is fixed at
birth, $S_i=c_Sx_ii^{-2/3}$. With a rate per unit area $\langle o\rangle/2A(t)$ on each side
and $A(t)=3c_St^{1/3}$,
\begin{equation}
  k_i=o_i+\frac{\langle o\rangle}{2}\,x_i\Big[\Big(\frac{t}{i}\Big)^{2/3}-1\Big].
  \label{eq:k3d}
\end{equation}
Differentiating gives $dk_i/dt=\tfrac23(k_i-o_i+\langle o\rangle x_i/2)/t$, which is
$\tfrac23(k_i-\langle o\rangle/2)/t$ in mean field ($o_i\to\langle o\rangle$, $x_i\to1$). Under (ii) a hit
splits a face but adds no area, whereas under (i) every hit adds a face that is as
likely to be hit as any other. The rate is still linear in $k$, now with a slope below
the incidence value and a negative constant, $b=2/3$ and $a=-\langle o\rangle/3$, in agreement with the sum
rule~\eqref{eq:sumrule}. With this attachment rate the rate equation~\eqref{eq:rate}
applies unchanged, with $q_k$ the distribution of the number of plates met at birth, and gives
$\gamma=5/2$. In $d$ dimensions the same count gives
\begin{equation}
  b=\frac{d-1}{d},\qquad a=-\frac{d-2}{d}\langle o\rangle,\qquad
  \gamma=\frac{2d-1}{d-1},
  \label{eq:bd}
\end{equation}
the exponent of random Apollonian networks in $d$ dimensions~\cite{kolossvary2016}.
Here $b$ is exact within mean field, because cell volumes scale exactly as $1/t$; the
constant $a$ carries the one number we cannot deduce, the mean $\langle o\rangle$, which
equals $\bar m$ away from the walls. The two counts
agree in two dimensions only because $m/(2+m)=(d-1)/d$ when $m=2$.

The measured kernel selects count (ii). With the same estimator as in two dimensions
(three realisations, $T=2\times10^6$), a linear fit for $k\le200$ gives $\kappa=-1.477+0.660\,k$. The deduced kernel, with $\langle o\rangle=4.538$,
is $-1.513+0.667\,k$ [Fig.~\ref{fig:degree}(a)].
The stationary solution of Eq.~\eqref{eq:rec} with these constants and the measured
$q_k$ matches the simulated distribution [Fig.~\ref{fig:degree}(b)]. Count (i) fails because the faces of a plate are not
typical. An old plate is tiled by faces of linear size $t^{-1/3}$, so its number of
faces grows as $t^{2/3}$, more slowly than the $t^{\langle o\rangle/(2+\langle o\rangle)}=t^{0.694}$
that uniform hitting per incidence requires. The faces of a fixed old plate grow as $t^{0.663\text{--}0.668}$. A surrogate that hits
incidences uniformly with the same birth degrees gives $0.688$--$0.690$, and one that
hits by area gives $0.658$--$0.661$. Per unit area the rate is nearly uniform: in units
of the mean it is $1.04$ for old plates and $0.93$--$0.95$ for young ones, a few-per-cent
correction of the same sign as in two dimensions.

Beyond mean field the picture is the same as in two dimensions. Equation~\eqref{eq:k3d}
predicts the hits received by each plate with ratio $1.015$ and correlation $0.99$. At
fixed age the hits scale with birth area with elasticity $0.94$--$0.95$. Hits in
$(i,2i]$ barely predict hits in $(2i,T]$ at fixed area (log--log slope
$0.025$--$0.030$, against $0.55$ in the incidence surrogate). The dispersion around
each plate's own prediction is $1.5$--$4$ times Poisson, against $100$--$380$ times in
the incidence surrogate. The maximum-likelihood exponent of all plates is
$2.48$--$2.51$ for $k_{\min}=80$--$640$ (standard errors $0.014$--$0.064$); the area and incidence surrogates, whose asymptotes are
$2.5$ and $2.44$, read $2.48$--$2.53$ and $2.40$--$2.46$ at the same cuts. Fits restricted
to plates away from the walls are biased upward ($2.67$--$3.18$), because the oldest,
largest plates touch the walls. Three-dimensional SR materials therefore have
$\gamma=5/2$. The host of a
new plate satisfies $h/t\to U^3$ (measured $P(h/t\le0.01)=0.217$, against $0.215$), so the
thickness ratio to the host is $U^{3\alpha}$.

A plate nucleated in free volume $F$ has area $S_t=c(t)(F/t)^{2/3}$, with
$c(t)\to c_S=1.23$ measured in zero-thickness runs, so the free volume obeys
\begin{equation}
  \begin{aligned}
  \frac{dF}{dt}&=-\lambda_0c(t)F^{2/3}t^{-\alpha-2/3},\\
  F_\infty^{1/3}&=1-\frac{\lambda_0}{3}\sum_tc(t)t^{-\alpha-2/3}.
  \end{aligned}
  \label{eq:F3d}
\end{equation}
It matches the simulated free volume to within $0.4\%$ in the tunable phase. The sum
converges only for $\alpha>1/3$, which places the jamming boundary at $\alpha=1/3$ in three
dimensions, and for $\alpha>1/3$ Eq.~\eqref{eq:F3d} predicts a second branch,
$\lambda_0^*\simeq3/[c_S\zeta(\alpha+2/3)]$, the analogue of the $\lambda_0$ branch in two
dimensions; we have not located it in simulations. For small $\lambda_0$ in the tunable
phase, $\rho\simeq\lambda_0c_S\zeta(\alpha+2/3)$.

Clustering is high because a new plate's section polygon is a cycle of plates already in
contact. Each newcomer closes $4.39$ triangles along its polygon and $2.52$ more through
contacts outside its cell. Each received link brings $q=3.26$ triangles. Then
$C(k)=2[e_0+q(k-o)]/[k(k-1)]\to6.5/k$, which agrees with the data within $1\%$ for $k\ge16$,
and $\langle C\rangle=0.72$. The three-dimensional random Apollonian network has the same
$\gamma$, $C(k)=6/k$ and $\langle C\rangle=0.82$. The mean neighbour degree decreases with
$k$ and grows as $T^{1/3}$.

Thickness and jamming change neither $\gamma$ ($2.49$--$2.55$) nor $C$ ($0.71$--$0.72$).
With three orthogonal orientations every cell is a box and every plate a rectangle
($m=4$). The two counting arguments then coincide at $b=2/3$, and $\gamma\simeq2.5$. These
boxes cluster ($C=0.55$), whereas rectangles in two dimensions do not. Beams, which need
a capture rule, are discussed among the open problems in Sec.~\ref{sec:conc}.

\section{Consequences for structure--property relations}
\label{sec:energy}

At fixed $\alpha$, the thickness of ligament $i$ is $\lambda_0i^{-\alpha}$ exactly,
and its degree and length both scale as $i^{-1/2}$ on average
[Eq.~\eqref{eq:kl}]. The marginal laws reported in
Ref.~\cite{both2026}, an energy that decays as $i^{-(3/4+\alpha/2)}$ with the
introduction time and the corresponding powers of $k$, $\ell$ and $\lambda$, are one
age law written in different variables. A product $U\sim k^u\lambda^v$ reproduces the
age law for every $\alpha$ only if $u/2+\alpha v=3/4+\alpha/2$, that is $u=3/2$ and
$v=1/2$, the reported exponents. Because $k\propto\ell$ at fixed age
[Eq.~\eqref{eq:kl}], the same data are described equally well by
$U\sim\ell^{3/2}\lambda^{1/2}f(\theta)$. Degree, length and age cannot be separated
from cohort scaling alone.

Separating them requires the per-ligament energies themselves: a within-cohort
regression of $U$ on $k$, $\ell$ and $\lambda$ would show whether degree adds anything
beyond length and age. In a slender-beam frame model (rigid T-junctions, $T\le5000$), which is not the
finite-element model of Ref.~\cite{both2026}, fits on degree and on length were
indistinguishable ($R^2=0.634$ and $0.631$).

The same reasoning applies to design. Thickness is a function of birth order, and
degree and length are statistically slaved to birth order, so $\alpha$ and $\lambda_0$
act on the material through the age of each ligament on a fixed network.
Connectivity becomes a design variable only when the rule is changed so that the hit
rate stops being proportional to length, as with coherent orientation fields, or when
the dimension changes $b$, the relation between the measure of a ligament and its birth order
(Secs.~\ref{sec:gamma} and \ref{sec:3d}).

\section{Conclusions}
\label{sec:conc}

The SR dual belongs to the class of local growth rules. Geometrically the construction
is Cowan's recursive division of cells~\cite{cowan2010}; as a graph each new node joins
two nodes on the boundary of one face. This single face-local step produces the
hierarchical clustering spectrum through corner cuts, the excess of shortest paths
through square-closing splits, and the degree law. The degree law follows from the
rate equation approach~\cite{krapivsky2000} with a kernel deduced from the rule:
a ligament of degree $k$ borders $k$ cells, so $b=2/\langle n\rangle=1/2$, the sum rule
gives $a=0$, and $p_k=12/[k(k+1)(k+2)]$ follows with no free parameter. The measured
kernel is $\kappa=0.50\,k$ and the predicted distribution is within $4\%$ of the
simulations. Preferential attachment enters only as this mean-field outcome. Beyond mean field, the recursion assigns each ligament its length
at birth, $\ell_i\propto i^{-1/2}$, and the degree is a Poisson count of hits with mean
proportional to length times remaining time, without reinforcement. In $d$ dimensions
the hierarchy fixes $b=(d-1)/d$, and plates in three dimensions have $\gamma=5/2$, with
the measured kernel $-1.48+0.66\,k$ against the deduced $-1.51+0.67\,k$. Keeping both references in view settles
which features are fixed by the class and which can be designed. The exponent
$\gamma=3$ (and $5/2$ for plates in three dimensions) and the equality $\langle n\rangle=4$
are fixed, and in mean field so are
the host-age law $h/t\to U^2$ and the stationarity of the thickness ratio. Clustering
is designable through the angle set and the selection weight. The degree exponent is
designable only through $b$: rules that make the hit rate grow faster than length, or a
change of dimension.

Several points remain open. The rate $2/L(t)$ per unit length is a mean-field rate:
old ligaments receive a few per cent more hits than it predicts ($2\%$ in two
dimensions, $4\%$ in three), which sets the tail amplitude; we have no theory of this correction. The per-hit triangle probability, which rises from $0.41$ to $0.57$ with the
degree, and the fraction $0.048$ of closures through a link off the face require the joint law of cell shapes
next to long ligaments. Under coherent orientation fields we observe
$\gamma\simeq2.4$--$2.5$ up to $T=10^6$; deducing $b$ from the length scalings gives the
direction of the change but not its value. In three dimensions the mean number of plates met by a new plate,
$\bar m=4.577$, has no closed form, and it enters the kernel through $a=-\langle o\rangle/3$. Beams need a capture rule. If a rod of radius $R_t=\lambda_t/2$ stops at first
contact, its mean free path is $\ell_t\sim1/\sum_j\ell_jR_j$; self-consistency gives
$\ell_t\sim t^{-(1-\alpha)/2}$, hits $\propto\ell_iR_i\propto i^{-(1+\alpha)/2}$ and, as a
conjecture, $\gamma=(3+\alpha)/(1+\alpha)$, so thickness would tune $\gamma$. Our
simulations up to $T=10^5$ are dominated by the walls, and the capture rule of
Ref.~\cite{both2026} is not specified in its main text. Finally, the growth of the shortest-path excess with $T$ is
consistent with both a log-quadratic and a weak power law over the range studied.

More broadly, the SR dual is a physical instance of preferential attachment as an
effective result rather than a mechanism, and the clustering that comes with the local sampling identifies the class.

\begin{acknowledgments}
All code is available at \url{https://github.com/av2atgh/scalerich}.
The calculations were assisted by Claude Opus~5, an AI assistant developed by Anthropic. Nodes \& Links Ltd
provided support in the form of salary for Alexei Vazquez but did not have any additional role in the conceptualization of the study, the
analysis, the decision to publish or the preparation of the manuscript.
\end{acknowledgments}

\bibliography{references}

\end{document}